\documentclass[english,12pt,a4paper]{article}
\usepackage[T1]{fontenc}
\usepackage[latin9]{inputenc}
\usepackage[letterpaper]{geometry}
\usepackage{array}
\usepackage{amsmath}
\usepackage{amssymb}
\usepackage{bbm}
\usepackage{graphicx}
\usepackage{subfigure}

\makeatletter

\providecommand{\tabularnewline}{\\}

\usepackage{youngtab}
\usepackage{cite}
\usepackage{a4wide}

\newcommand{\fund}{\ensuremath{\Box}}
\newcommand{\afund}{\ensuremath{\overline{\Box}}}

\newcommand{\SU}[1]{\mathrm{SU}(#1)}
\newcommand{\U}[1]{\mathrm{U}(#1)}

\newcommand{\vev}[1]{\langle #1 \rangle}

\newcommand{\sunc}{\mathrm{SU(N)}}
\newcommand{\sun}{\mathrm{SU(n)}}
\newcommand{\Uo}{\mathrm{U(1)}}
\newcommand{\miss}{\mu_{\mbox{\tiny ISS}}}
\newcommand{\missp}{\mu_{\mbox{\tiny ISS}'}}
\newcommand{\mpl}{\mu_{\mbox{\tiny UV}}}
\newcommand{\mir}{\mu_{\mbox{\tiny IR}}}
\newcommand{\wir}{W_{\mbox{\tiny IR}}}
\newcommand{\kir}{K_{\mbox{\tiny IR}}}
\newcommand{\wuv}{W_{\mbox{\tiny UV}}}
\newcommand{\sunf}{\mathrm{SU(F})}
\newcommand{\nc}{\mathrm{N}}
\newcommand{\nfc}{\mathrm{\breve{F}}}
\newcommand{\nnc}{\mathrm{n}}
\newcommand{\nf}{{\mathrm{F}}}

\newcommand{\mm}{{\phi}}

\newcommand{\pp}{\varphi}
\newcommand{\p}{\Phi}

\newcommand{\beq}{\begin{equation}}
\newcommand{\eeq}{\end{equation}}

\newcommand{\tr}{\operatorname{tr}}

\makeatother

\usepackage{babel}

\begin{document}

\begin{titlepage}

{\flushleft{IPPP/10/17 \\DCPT/10/34}}\\[2.0cm]

\begin{center}

\textsc{ \LARGE Holographic Metastability}\\[1.5cm]

{ Steven Abel\footnote{s.a.abel@durham.ac.uk} and Felix Br\"ummer\footnote{felix.bruemmer@durham.ac.uk}\\
\vspace{1ex}
\emph{Institute for Particle Physics Phenomenology}\\
\emph{Durham University, DH1 3LE, UK}}

{\emph \today}

\begin{abstract}
\noindent We show how supersymmetric QCD in a slice of AdS can 
naturally acquire metastable vacua. 
The formulation closely follows that of  
Intriligator, Seiberg and Shih (ISS), with an ``electric'' sector on the UV brane and a 
``magnetic'' sector on the IR brane. However the 't~Hooft anomaly matching that 
constrains the Seiberg duality central to ISS is replaced by anomaly inflow and 
cancellation, and the source of strong coupling is the CFT to which the theory couples 
rather than the gauge groups. The theory contains an anomaly free $R$-symmetry that, when 
broken by UV effects, leads to an O'Raifeartaigh model on the IR brane. In contrast to 
ISS, the $R$-symmetry breaking in the UV can be maximal, and yet the $R$-symmetry 
breaking in the IR theory remains under strict control: there is no need for 
retrofitting of small parameters. 

\end{abstract}

\end{center}
 
\end{titlepage}

\section{Introduction and Conclusion}

Strong coupling will surely be central to our eventual understanding of how 
supersymmetry (SUSY) is broken in nature, as was suggested in the pioneering  
work of Refs.\cite{Witten:1981nf,Affleck:1983rr,Affleck:1983mk,ADS:susybreaking}. More recently 
Seiberg's proposal of electric/magnetic duality in ${\cal N}=1$ theories 
\cite{S:Duality,Intriligator:1995au}  famously found application to SUSY breaking
in the seminal discovery by Intriligator, Seiberg and Shih (ISS) of metastable vacua in the free-magnetic phase of supersymmetric
QCD \cite{ISS}. Although metastability had appeared in the earlier literature in various guises
(for example Refs.\cite{Ellis:1982vi,Dine:1995ag,Dimopoulos:1997ww,Luty:1998vr,Banks:2005df}), the surprise was 
that it automatically arose in virtually the simplest model that one can write down. 
That work stimulated further efforts, both in direct model-building applications, and more generally 
in understanding the general role of metastability in SUSY breaking, and its mediation to the Standard Model. 

Given this considerable advance, surprisingly little has been said about metastability using the other well known 
tool for dealing with strong coupling, namely the AdS/CFT correspondence \cite{Maldacena:1997re,Gubser:1998bc,Witten:1998qj}. 
Therefore the purpose of this paper is to provide a 
concrete framework for constructing metastable models in the holographic framework. 
The model we shall present is guided fairly rigidly by the ISS model itself, but rendered 
so as to fit in a slice of AdS$_5$. The ultra-violet (UV) brane (i.e. the fundamental sector) contains 
a 4D ${\cal N}=1$ theory resembling the electric formulation of SQCD, and the infra-red (IR) brane 
contains a theory resembling the magnetic formulation. The bulk contains the gauged flavour symmetries 
of SQCD, and the constraints of anomaly cancellation of those symmetries replace the 
't~Hooft anomaly matching conditions of Seiberg duality.
Other familiar features of Seiberg duality, 
such as baryon matching, also have an equivalent realisation in the holographic models we present.
Note however that this is {\em not} Seiberg duality: the strong coupling is taking place in the CFT theory 
to which the weakly gauged bulk theory couples.

As in the ISS model (and indeed all metastable models \cite{Nelson:1993nf}) $R$-symmetry will play an important role. 
The original theory has an $R$-symmetry that is exact and anomaly free. If the $R$-symmetry 
is broken spontaneously and maximally in the UV theory, then the warping, 
together with the (gauged) flavour symmetries, greatly constrain how it can appear in the IR theory.
The IR theory closely resembles the ISS model, and metastable SUSY breaking ensues. The 
SUSY restoring minima may be entirely contained within the perturbative 4D low energy description for a specific choice of 
flavours and colours (the same choice in fact that would put one in the free magnetic phase of SQCD
if one were doing Seiberg duality proper). 

There are several benefits of using a holographic approach in this context, some of which were suggested in 
Refs.\cite{Dudas:2007hq,Abe:2007ki}. As already mentioned, the UV theory can maximally break $R$-symmetry, 
but the theory on the  IR brane maintains an approximate $R$-symmetry. This contrasts with the 
ISS model where the $R$-symmetry breaking term is a tiny mass deformation \cite{ISS}: to generate dynamically (i.e. retrofit \cite{Dine:2006gm,Dine:2006xt}) 
such a term requires another strongly coupled sector. In the holographic approach on the other hand 
one is effectively retrofitting using the same strong coupling that leads to dynamical SUSY breaking. 

The paper is organised as follows: in the following section we recapitulate, for the purposes of 
comparison, the model of ISS. Following that we introduce in section 3 an equivalent holographic 
model that closely mimics the electric and magnetic phases of Seiberg duality that are central to ISS. 
The main difference is that the holographic model 
is constrained by anomaly cancellation rather than by anomaly matching as mentioned 
above; thus we spend some time discussing how anomaly cancellation and in 
particular anomaly inflow works for the 5D formulation. We also make other connections to 
standard Seiberg duality, for example in baryon matching. Section 4 then introduces a deformation 
to break SUSY dynamically. The deformation in question is one that maximally breaks
$R$-symmetry. The result on the IR brane is a retrofitted O'Raifeartaigh 
model of the ISS type, with the SUSY breaking parameter being exponentially suppressed by 
the warping. As in ISS the SUSY breaking is metastable, with SUSY restoring 
minima appearing in the low energy theory due to the anomalous nature of the remaining $R$-symmetry in that sector. 
The holographic configuration and spontaneous $R$-breaking ensures that any other $R$-violating operators are 
even more suppressed and that the metastability is therefore preserved. 

\section{Metastability in Seiberg duality (ISS)}

Our aim is to transpose the dynamical SUSY breaking properties of
strongly coupled SQCD into a holographic configuration. Before doing the latter we
should first review the former. In particular it is worth re-examining
the special role that the global symmetries play in these theories.
(In our AdS incarnation of ISS, these symmetries will be gauged.)
ISS examined the IR free magnetic dual of an asymptotically free $\sunc$
theory with $\nf$ flavours \cite{ISS}. With an empty superpotential this theory
has a global $\sunf_{L}\times\sunf_{R}\times\Uo_{B}\times\Uo_{R}$
symmetry. These global symmetries are anomaly free with respect to
the gauge symmetry. There is also an anomalous $\Uo_{A}$ symmetry
which (since it cannot be consistently gauged) will be irrelevant
for our discussion. The particle content is shown in Table \ref{sqcd0}.
\begin{table}
\centering{}\begin{tabular}{|c||c|c|c|c|c|}
\hline 
$ $ &  $\SU\nc$ & $\sunf_L$ & $\sunf_R$ & $\U1_{B}$ & $\U1_{R}$\tabularnewline
\hline
\hline 
$Q$ & $\fund$ & $\fund$ & 1 & $\frac{1}{\nc}$ & $1-\frac{\nc}{\nf}$\tabularnewline
\hline 
$\tilde{Q}$ &$\afund $& 1 & $\afund$ & $-\frac{1}{\nc}$ & $1-\frac{\nc}{\nf}$\tabularnewline
\hline
\end{tabular}\caption{\emph{Spectrum and anomaly free charges in }SQCD\emph{.}\label{sqcd0}}

\end{table}

The magnetic dual theory (which we refer to as $\overline{\mbox{SQCD}}$)
has a gauged $\SU\nnc$ symmetry, where $\nnc=\nf-\nc$ \cite{S:Duality,Intriligator:1995au}.
Its spectrum is given in Table \ref{sqcd0-1}. (Throughout we will denote
magnetic superfields with small letters and electric superfields with
capitals.)
\begin{table}
\centering{}\begin{tabular}{|c||c|c|c|c|c|}
\hline 
$ $ &  $\SU\nnc$ & $\sunf_L$ & $\sunf_R$ & $\U1_{B}$ & $\U1_{R}$\tabularnewline
\hline
\hline 
$q$ & $\fund$ & $\afund$ & 1 & $\frac{1}{\nnc}$ & $1-\frac{\nnc}{\nf}$\tabularnewline
\hline 
$\tilde{q}$ & $\afund$& 1 & $\fund$ & $-\frac{1}{\nnc}$ & $1-\frac{\nnc}{\nf}$\tabularnewline
\hline 
$\mm\equiv Q\tilde{Q}$ &1& $\fund$ & $\afund$ & 0 & $2\frac{\nnc}{\nf}$\tabularnewline
\hline
\end{tabular}\caption{\emph{Spectrum and anomaly free charges in} $\overline{\mbox{SQCD}}$\label{sqcd0-1}}

\end{table}
The two theories satisfy all the usual tests of anomaly and baryon
matching if one adds a superpotential \begin{equation}
\wir=\frac{1}{\mu}q\mm\tilde{q}.\end{equation}
The equation of motion of the elementary meson then projects the superfluous
composite meson $q\tilde{q}$ out of the moduli space of the magnetic
theory. By definition $\mu$ relates the vev of the dimension-two
composite meson $\mm=Q\tilde{Q}$ to the masses of the magnetic quarks
(i.e. $\mm/\mu$); it connects the dynamical scales of the two theories
as \begin{equation}
\Lambda^{b}\bar{\Lambda}^{\bar{b}}=(-1)^{\nf-\nc}\mu^{b+\bar{b}}\end{equation}
where $b$ and $\bar{b}$ are the SQCD beta function coefficients of the magnetic
and electric theories ($3\nc-\nf$ and $3\nnc-\nf$ respectively)
and where $\Lambda$ and $\bar{\Lambda}$ are their respective dynamical
transmutation scales. Note that in this
expression the quarks are assumed to be canonically normalized, but
$\mm$ needs normalizing: generally its K\"ahler potential will have
a leading term \begin{equation}
\kir=\frac{\mm\mm^{\dagger}}{h^{2}\mu^{2}},\end{equation}
where $h$ is some constant expected to be of order unity. One can
define a normalized meson for the magnetic theory, $\pp=\mm/(h\mu)$,
so that the K\"ahler potential is canonical, \begin{equation}
\kir=\pp\pp^{\dagger},\end{equation}
and then the superpotential has an unknown Yukawa coupling \begin{equation}
\wir=h\, q\pp\tilde{q}.\end{equation}
In either case, if the coupling of the electric theory (and hence
$\Lambda$) is known, then there are two unknown parameters, $h$
and $\mu$, defining a class of Seiberg duals. 

Now, the observation of Ref.\cite{ISS} was that if one adds a mass term
to the electric quark \begin{equation}
\wuv=m_{Q}\,Q\tilde{Q}\end{equation}
then the classical superpotential of the magnetic theory is of the O'Raifeartaigh
type: \begin{equation}
\wir=h\, (q\pp\tilde{q}-\miss^{2}\, \pp)\end{equation}
where $\miss^{2}=\mu\,m_{Q}$. For $\nf>\nc$ the so-called rank
condition implies that supersymmetry is broken; that is \begin{equation}
F_{\pp_{j}^{i}}=\tilde{q}^{j}\cdot q_{i}-\miss^{2}\,\delta_{i}^j=0\end{equation}
can only be satisfied for a rank-$\nnc$ submatrix of the $F_{\pp}$.
The height of the potential at the metastable minimum is then given
by \begin{equation}
V_{+}(0)=\nc\,|h^{2}\miss^{4}|\, .\end{equation}
The supersymmetric minima in the magnetic theory are located by allowing
$\pp$ to develop a vev. The $q$ and $\tilde{q}$ fields
acquire masses of $\langle h\pp\rangle$ and can be integrated out,
whereupon one recovers a pure $\sun$ Yang-Mills theory with a nonperturbative
contribution to the superpotential of the form \begin{equation}
\wir^{\mbox{\tiny (dyn)}}=\nnc\left(\frac{h^{\nf}\mbox{det}_{\nf}\pp}{\bar{\Lambda}^{\nf-3\nnc}}\right)^{\frac{1}{\nnc}}.\end{equation}
This leads to $\nc$ nonperturbatively generated SUSY preserving minima
at \begin{equation}
\label{phivev}
\langle h\pp_{i}^{j}\rangle=\miss\,\epsilon^{-(\frac{3\nc-2\nf}{\nc})}\delta_{i}^{j}\end{equation}
where $\epsilon=\miss/\bar{\Lambda}$, in accord with the Witten
index theorem. The minima can be made to appear far from the origin
if $\epsilon$ is small and $3\nc>2\nf$, the condition for the magnetic
theory to be IR-free. The positions of the minima are bounded by the
Landau pole such that they are always in the region of validity of the macroscopic
theory.

It is interesting to note (and obliquely relevant for what comes later) that one can find a 
whole class of electric theories that flow to the same IR physics, by performing 
multiple dualities. Dualizing the magnetic theory again one finds an electric theory with two singlet ``mesons'', 
$\mm$ and $\p$, say, the latter having the same quantum numbers as the magnetic composite 
meson $q\tilde{q}$. The electric superpotential is then 
\beq
\wuv = -\frac{1}{\mu}  Q \p \tilde{Q}+\frac{1}{\mu }  \mm  \;\!\p - m_Q \mm\, .
\eeq
One can then integrate out $\p$ and $\mm$ whereupon one recovers the 
original electric theory, the standard dual-of-a-dual test of Seiberg duality. Or one 
can keep the mesons in the model (choosing parameters
such that their masses are below the strong coupling scale of the electric theory). 
Upon dualizing {\em again} one finds a magnetic model with three mesons that 
has an ISS-like metastable minimum as before, but with SUSY breaking distributed 
equally between the magnetic mesons. In fact the mass $m_Q$ can also be 
arbitrarily divided between $\mm$ and $Q\tilde{Q}$.  By continued dualizing  
any number of mesons can be introduced.

Now, an important role is played by the $R$-symmetry of the model.
The mass term $m_{Q}Q\tilde{Q}$ explicitly breaks the anomaly-free
$R$-symmetry but leaves behind an anomalous $R$-symmetry which is
a linear combination of the $\Uo_{R}$ in Table \ref{sqcd0} and the
orthogonal anomalous $\Uo_{A}$ (which we do not display - see Ref.\cite{Abel:2007nr}
for more details). It is the anomalous nature of \emph{this} symmetry
which (in accord with the Nelson--Seiberg theorem \cite{Nelson:1993nf})
allows the supersymmetric minima to appear, but in a controlled manner.
Given the anomalous nature of the remaining $R$-symmetry one is then
entitled to add further operators to the electric theory in order
to do phenomenology (e.g. gauge mediation with non-zero gaugino
masses). However the attractive feature of this set-up is that, when
such operators are translated to the magnetic theory, factors of $\mu/\mpl$
(where $\mpl$ is the fundamental scale of new physics in the UV
theory) are induced. Consequently the approximate $R$-symmetry of
the magnetic theory remains and the metastability is left intact.
This was the main point of ref.\cite{MN} which showed how to exploit
the set-up to do very simple standard gauge mediation. To illustrate
the point (without having to discuss mediation), consider adding the
operator \begin{equation}
\wuv\supset\frac{(Q\tilde{Q})^2}{\mpl}\end{equation}
to the electric theory. In the magnetic theory this becomes a very
small mass term, \begin{equation}
\wir\supset\frac{\mu^{2}}{\mpl} \pp^{2},\end{equation}
which introduces new minima at vevs of order $\pp\sim \mpl\left(\frac{\miss^{2}}{\mu^{2}}\right)$
that can easily be made larger than $\mu$. One of the reasons that
the ISS set-up is useful for phenomenology is therefore that, as well
as generating a linear term in the magnetic theory, the mass deformation
operator $\tilde{Q}Q$ is the lowest dimension operator in the electric
theory that one can write down. The obvious question is what suppresses
$m_{Q}$ itself. Indeed, one requires $\miss\ll\mu$ (or equivalently
$m_{Q}\ll\mu$) in order to have long lived metastable vacua in the
magnetic theory: then $m_{Q}$ is smaller than any scales that
naturally appear in the electric theory. Clearly in order to achieve
this some extension of the model is required in order to retrofit
this parameter dynamically \cite{Dine:2006gm,Dine:2006xt}. This could
for example be a third sector (besides the SUSY breaking and the
mediating sectors) that becomes strongly coupled at a scale $\Lambda'$
and generates mass parameters of order $m_{Q}\sim(\Lambda')^{3}/\mpl^{2}$
\cite{Aharony:2006my} or $m_{Q}\sim\Lambda'$ \cite{Brummer:2007ns}.

Let us summarize the important features of ISS which we wish to reproduce
in the holographic context:
\begin{enumerate}
\item The theory consists of a UV phase and an IR phase controlled by large
global symmetries. 
\item The theory has an anomaly free $R$-symmetry that is broken by a deformation
of the UV theory.
\item The deformation induces metastable minima in the IR theory that are
protected by a remaining anomalous $R$-symmetry.
\item The SUSY breaking is under control. In particular the composite nature
of the IR theory means that any additional deformations one adds to
the fundamental UV theory are unable to destabilize its metastable
vacua.
\end{enumerate}
These points are of course all positive. On the down side we can add
the feature that the original deformation is unnaturally small and
has to be retrofitted. It turns out that we will be able to avoid
this latter problem altogether in a holographic set-up. 

Before turning to the specifics of our model, we should briefly comment on the relation of our work to Ref.\cite{Dudas:2007hq}
which also considered metastability in a holographic set-up. That work discussed tunnelling in a theory which was equivalent 
to the magnetic theory of ISS, but with the superpotential terms split between the IR and UV branes (e.g. $\wuv=h\,q \pp\tilde{q}$ on the 
UV brane and $\wir=\miss^2 \pp $ on the IR brane). As such, the questions above
(i.e. the underlying $R$-symmetry, and the method of its breaking) are outside the scope of that framework, and 
are ultimately on  the same 
footing as in the ISS model itself. For example, given that the $R$-symmetry is anomalous there is no reason not to also 
include a $\mir\pp^2$ term on the IR brane, where $\mir$ is the warped-down mass scale on the IR brane. 
This would destabilize the hierarchy by introducing a new global supersymmetric minimum at 
$\pp \sim \miss^2 /\mir$. Moreover, the quarks acquire vevs of order $\miss$ in the metastable minimum,
and to be able to ignore the effect of their coupling to KK modes one probably requires $\miss<\mir$, which implies
that this second global SUSY minimum lies close to the origin in $\pp$.
 The way to prevent this happening without fine-tuning is to appeal to an 
underlying $R$-symmetry which is approximately maintained because of the composite nature of the IR theory and some underlying 
dynamics. But for this one would need information about the electric Seiberg dual theory which is not included in that discussion.

\section{A holographic rendering of Seiberg duality}

In this section we begin to outline the scheme
for translating ISS metastability to a holographic set-up. 
We should stress at the outset than the model is {\em not} Seiberg duality, but 
reproduces the defining features that were important to the ISS model.
In particular the strong coupling is not in the explicit gauge groups, but is in the 
CFT to which they couple. 
Nevertheless many of the characteristics of the model closely resemble those of Seiberg duality. 
It will be ${\cal N}=1$ supersymmetric, with an $\sunc$ UV theory and an $\SU\nnc$ IR theory, 
and it will preserve a global $R$-symmetry. In the following section we will then 
show how to break the symmetries to get metastable SUSY breaking on the IR brane.

We will work in the RS1 scenario \cite{Randall:1999ee}, compactifying on an $S^{1}/(Z_{2}\times Z_2')$ interval with branes
at $y=0$ and $y=\pi R$, where $y$ denotes the fifth dimension.
We use the AdS$_5$ metric \begin{equation}
ds^{2}=e^{-2ky}\eta_{\mu\nu}dx^{\mu}dx^{\nu}-dy^{2}.\end{equation}

Now, one feature of Seiberg duality is that many aspects (such
as baryon matching as we shall see later) resemble what one would have in a model
where a unified colour $\sunf$ is broken to $\sun\times\sunc$.
Guided by this, we shall take the bulk gauge symmetry to be this group times the 
non-abelian flavour symmetries $\sunf_{L}\times\sunf_{R}$.
In order to distinguish the $\SU\nf$'s we shall put a breve over
the one that splits into colours, $\SU\nfc$. Thus our
total bulk gauge symmetry is $\SU\nfc\times\sunf_{L}\times\sunf_{R}$.
The $\SU\nfc$ is to be broken to $\sun\times\sunc$ on both
branes by orbifold boundary conditions, and the $\sunf_L\times\sunf_R$ will be broken to its diagonal subgroup by 
bulk field vacuum expectation values.
Guided by Seiberg duality, we require that the model has to separate
$\SU\nc$ quarks on the UV brane from $\SU{\nnc}$ quarks on the IR
brane. An important check for Seiberg duality 
is  't~Hooft anomaly matching, the fact that the flavour 
anomalies of both electric and magnetic theories are the same. Anomalies
play a similar role in determining our holographic set-up
because the flavour symmetries are all gauged in the bulk. Specifically, in order to 
achieve anomaly cancellation, one can simply put the quarks of the electric theory
on the UV brane and the quarks and mesons of the magnetic theory, but with all their 
flavour charges conjugated, 
on the IR brane. The bulk we shall assume to be empty of any $\SU\nfc$ charged matter
and vector-like with respect to $\sunf_{L}\times\sunf_{R}$. 
This ensures that the gauge anomalies cancel. Maintaining an anomaly free $R$-symmetry 
then requires only a modest amount of finessing. 

\begin{figure}[ht]
\begin{centering}
\includegraphics[angle=0,scale=0.15]{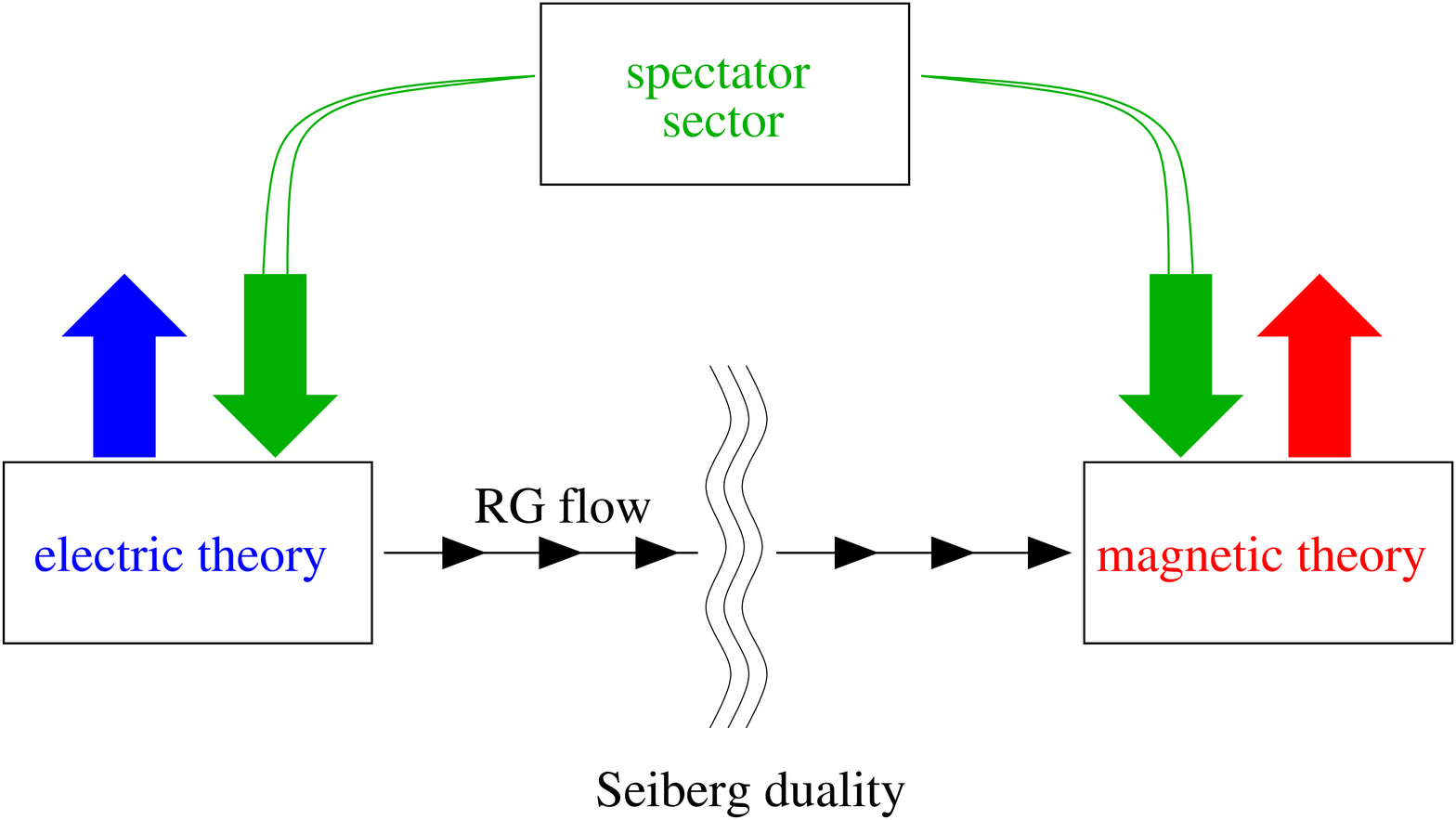}
\includegraphics[angle=0,scale=0.15]{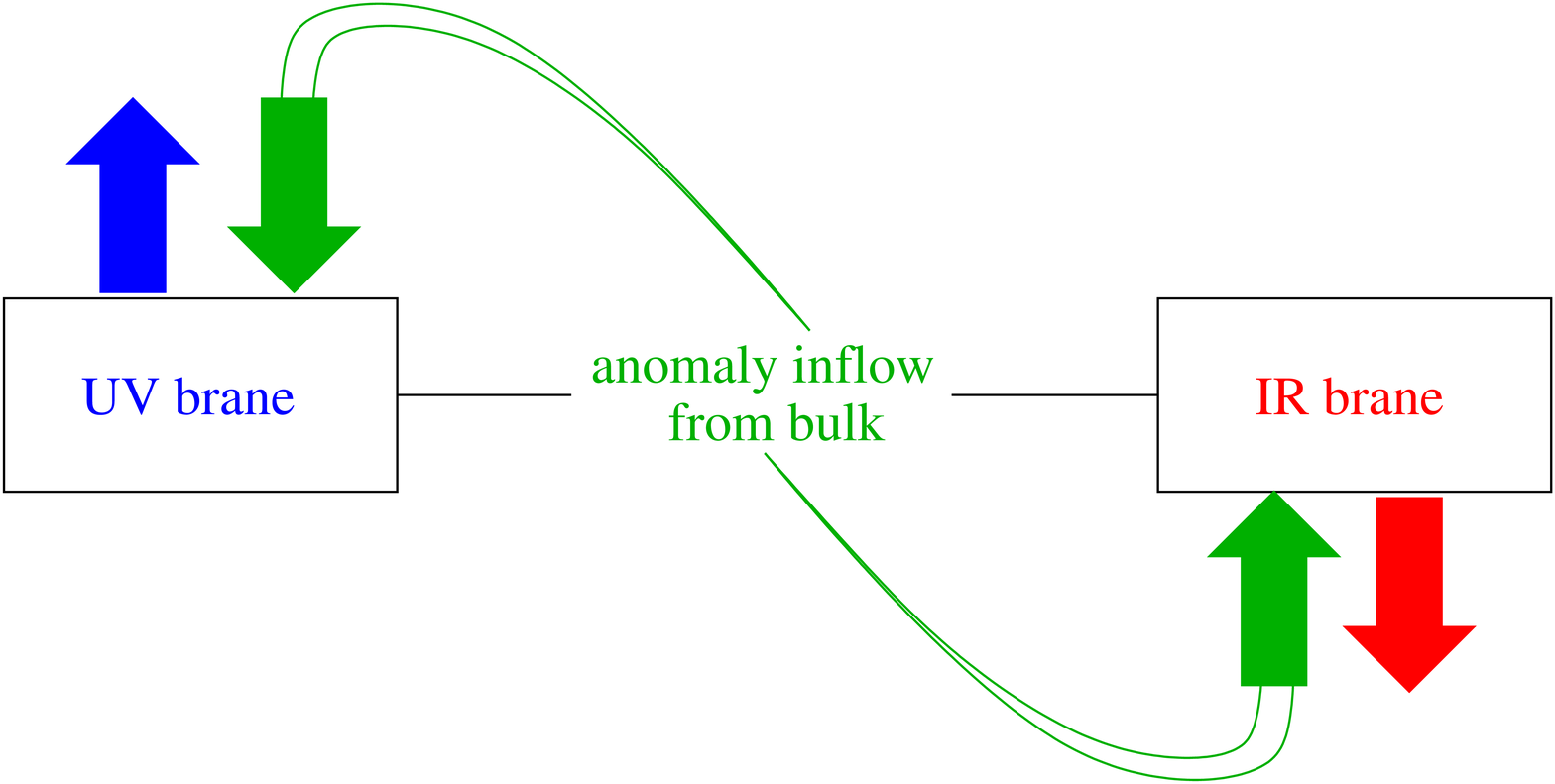}
\par\end{centering}
\caption{\it On the left, the schematic set-up for Seiberg duality (and ISS metastability). 
The blue and red arrows indicate the flavour anomalies generated by the electric and magnetic theories, which
may be cancelled by adding a ``spectator sector'' in the 't~Hooft anomaly matching procedure.
On the right, the set-up for holographic metastability. The flavour charges of one sector are conjugated
with respect to Seiberg duality so that the gauge anomalies on the IR brane (in red) are now opposite to those
of the UV sector (in blue). Both are cancelled by anomaly inflow from a Chern--Simons term in the bulk,
which gives equal and opposite contributions to anomalies on the branes.} 
\label{fig:anomaly}
\end{figure}

The picture is as shown in Figure~\ref{fig:anomaly}. On the left, the schematic set-up for 
Seiberg duality (and ISS metastability). In 't~Hooft anomaly matching, one imagines gauging 
flavour, which requires all anomalies to be cancelled by some additional spectator sector 
(uncharged under the colour gauge group). The new sector remains 
unchanged when the colour gauge group becomes strongly coupled, hence the magnetic theory's flavour 
anomalies should be the same. On the right, the set-up for holographic metastability. 
The flavour charges of one sector are conjugated, such that flavour anomalies of one theory
can cancel those of the other (on the level of the 4D effective theory); or 
in the language of the 5D theory, 
such that the anomalies on the IR brane and those on the UV brane can be cancelled 
simultaneously by anomaly inflow from the bulk 

\begin{figure}[ht]
\begin{centering}
\includegraphics[angle=0,scale=0.32]{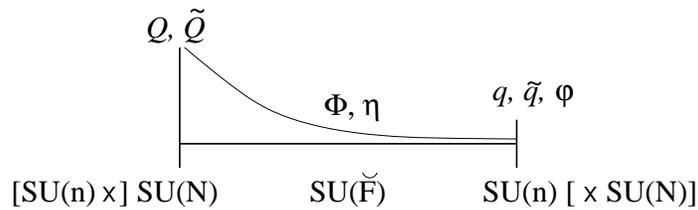}
\par\end{centering}
\caption{\it The field content of the model. The $\SU{\nfc=\nc+\nnc}$ bulk gauge group is
broken to $\SU\nnc\times\SU\nc$ on the branes. On the UV brane the ``electric quarks'' 
$Q,\tilde Q$ transform only under $\SU\nc$ and are singlets under $\SU\nnc$. The 
opposite is true for the ``magnetic quarks'' $q,\tilde q$ on the IR brane. There are gauge
singlets $\Phi$ and $\eta$ in the bulk, and $\varphi$ on the IR brane. In addition there
is a weakly gauged $\SU{\nf}_L\times\SU{\nf}_R$ flavour symmetry.}
\label{fig:model}
\end{figure}

This is the heuristic picture. Let us now present  the model in detail and come back to deal with 
anomalies and anomaly inflow more carefully in a moment.
To mimic the electric and magnetic phases of Seiberg duality, we will use the simple expedient of 
placing the relevant quarks on their respective branes, but will put some additional meson fields
in the bulk. As stated above the latter are uncharged with respect to $\SU\nc$ and $\SU\nnc$. 
The theory is given by Table \ref{sqcd0-3}, with
the separation into UV brane content, bulk and IR brane content indicated,
and depicted once more in Figure~\ref{fig:model}.
As well as the nonabelian symmetries already mentioned, there  are 
two additional abelian symmetries of relevance to us, $\Uo_B$ and $\Uo_R$, both of which are global.
As promised, although the assignment superficially resembles that of
Seiberg duality in Tables \ref{sqcd0} and \ref{sqcd0-1}, the $\sunf_{L}\times\sunf_{R}\times \Uo_B$
charges (i.e. everything that would be termed a flavour charge in
standard Seiberg duality) are reversed. \footnote{One could try to mimic more closely the 't~Hooft anomaly matching
of Seiberg duality by adding a sector uncharged under $\SU\nfc$ which
cancelled both UV and IR contributions to anomalies in exactly the
same way -- we do not think that would add to the discussion so we
will not explore the possibility here.}
\begin{table}
\centering{}\begin{tabular}{|c|c|c|c|c|c|}
\hline 
\multicolumn{1}{|c|}{$ $} & $G$ & $\sunf_{L}$ & $\sunf_{R}$ & $\U1_{B}$ & $\U1_{R}$\tabularnewline
\hline 
\hline 
$Q$ & $\fund$ & $\fund$ & 1 & $\frac{1}{\nc}$ & $1-\frac{\nc}{\nf}$\tabularnewline
\hline 
$\tilde{Q}$ & $\afund$ & 1 & $\afund$ & $-\frac{1}{\nc}$ & $1-\frac{\nc}{\nf}$\tabularnewline
\hline
\hline 
\multicolumn{1}{|c|}{$\p$} & 1 & $\afund$ & $\fund$ & 0 & $-2\frac{\nc\nnc}{\nf^{2}}$\tabularnewline
\hline 
{$\p^c$} & 1 & $\fund$ & $\afund$ & 0 & $2+2\frac{\nc\nnc}{\nf^2}$\tabularnewline
\hline 
$\eta$ & 1 & $\fund$ & $\afund$ & 0 & $2\frac{\nc}{\nf}$\tabularnewline
\hline 
$\eta^{c}$ & 1 & $\afund$ & $\fund$ & 0 & $2\frac{\nnc}{\nf}$\tabularnewline
\hline
\hline 
$q$ & $\fund$ & $\fund$ & 1 & $-\frac{1}{\nnc}$ & $1-\frac{\nnc}{\nf}$\tabularnewline
\hline 
$\tilde{q}$ & $\afund$ & 1 & $\afund$ & $\frac{1}{\nnc}$ & $1-\frac{\nnc}{\nf}$\tabularnewline
\hline 
$\pp$ & 1 & $\afund$ & $\fund$ & $0$ & $2\frac{\nnc}{\nf}$\tabularnewline
\hline
\end{tabular}\caption{\emph{Spectrum and anomaly free (with respect to the gauge groups)
baryon and R-charges on the UV brane, in the bulk, and on the IR brane.
The gauge group $G$ refers to the $\sunc$ factor in $\sun\times\sunc$ on the UV brane, 
to the $\sun$ factor on the IR brane.}
\label{sqcd0-3}}
\end{table}

The bulk fields $\eta$ and $\eta^c$ are conjugate 
pairs of $D=4$, $\mathcal{N}=1$ superfields which together constitute a 5D hypermultiplet.
We take $\eta$ to be even and $\eta^c$ to be odd under both orbifold projections.
The 5D action reads \cite{Gherghetta:2000qt, Marti:2001iw,Gherghetta:2006ha}
\beq
\begin{split}
{\cal S} ~=~&\int d^{5}x\Bigl\{\int d^{4}\theta\, e^{-2ky}\,\left[\eta e^{-V}\eta^\dagger +\eta^{c}e^{V}\eta^{c\dagger}\right]\\
&+\int d^{2}\theta e^{-3ky}\,\Bigl[ \eta^{c}\left( \mathcal{D}_{y}-\left( \frac{3}{2}-c_\eta\right)k\right)\eta
+\delta(y)\wuv+\delta(y-\pi R)e^{-3ky}\wir\Bigr]+{\rm h.c.}\Bigr\}
\end{split}
\eeq
where $\mathcal{D}$ is the covariant derivative and where $c_{\eta}$ leads to a 
bulk Dirac mass. We assume that no components of the gauge fields get vevs, so we
may for this discussion set $\mathcal{D}_{y}\equiv\partial_{y}$. Since the brane superpotentials
$\wuv$ and $\wir$ cannot depend on the odd fields $\eta^c$, the $F$-term equations are
\begin{eqnarray}
e^{-2ky}F^\dagger_{\eta} & = & e^{-3ky}\left(\partial_{y}-\left(\frac{3}{2}+c_\eta\right)k\right)\eta^{c}-\delta(y)\partial_{\eta}\wuv-\delta(y-\pi R)e^{-3k\pi R}\partial_{\eta}\wir\,,\nonumber \\
e^{-2ky}F^\dagger_{\eta^{c}} & = & e^{-3ky}\left(-\partial_{y}+\left(\frac{3}{2}-c_{\eta}\right)k\right)\eta\label{eq:f-terms}\, .
\end{eqnarray}
The bulk solution for $\eta$ is of the form \begin{equation}
\label{etavev}
\eta=A\, e^{(\frac{3}{2}-c_{\eta})k|y|}.\end{equation}
The normalized modes are $\hat{\eta}=e^{-ky}\eta$, so that if
$c_\eta>1/2$ we have localization around $y=0$ whereas 
$c_\eta<1/2$ gives localization around $y=\pi R$. These general solutions have to be matched 
to whatever vev $\eta$ may acquire due to brane interaction terms. The discussion for the other pair of bulk fields, 
$\Phi$ and $\Phi^c$, is entirely analogous.

The charges and global symmetries do not allow a superpotential on the UV brane (i.e. $\wuv=0$), but they do allow arbitrary Dirac masses in the bulk, and an IR superpotential of 
\beq
\label{wir}
e^{-{3 k \pi R}} \wir=h \, (q\pp \tilde{q}) + \mir\, (\hat{\eta} \pp)\, .
\eeq
where $\mir\sim e^{-k \pi R}\mpl$. (Note that $q$, $\tilde{q}$, $\pp$ and $\hat{\eta}$ are all defined to be the 
canonically normalized 4D fields.)

Finally in order to break the gauge group itself (without an adjoint Higgs) we can
give boundary conditions of \beq
V=\left(\begin{array}{c|c}
++ & +-\\ \hline
+- & ++\end{array}\right)\,\,;\,\,\Sigma=\left(\begin{array}{c|c}
-- & -+\\ \hline
-+ & --\end{array}\right),\eeq
to the 4D vector and 4D chiral components of the 5D vector supermultiplet, 
where the blocks refer to $\SU\nc$ and $\SU\nnc$ gauge groups. Note
that there are no remaining $\Sigma$ zero modes. 

\subsection{Anomalies and anomaly inflow}

In order consistently to gauge the flavour symmetries $\sunf_L$ and $\sunf_R$,
they should be anomaly free. Since anomalies are due to fermion zero modes,
which are localized in different regions of the internal space in our 
model, we briefly return to the issue of how local anomaly cancellation can be
guaranteed. First we should say that it is well-known that anomaly cancellation in the 4D effective 
theory is sufficient for cancelling any gauge anomalies 
in the 5D theory by a suitable bulk Chern--Simons term 
\cite{ArkaniHamed:2001is,Scrucca:2001eb,Barbieri:2002ic,GrootNibbelink:2002qp}. 
We will now briefly describe how this works in our case.

In five non-compact dimensions there is no anomalous divergence of a
classically conserved current, as there are no chiral fermions. 
By locality, anomalies on orbifolds can then only appear on the branes.
Both brane-localized fermions and bulk fermion zero modes can contribute. 
We define the sourceless generating functional $W[A]$ by
\begin{equation}
e^{i W[A]}=\int{\cal D}\Phi\; e^{iS[A,\Phi]},
\end{equation}
where the path integral is over all fields $\Phi$ except the gauge field $A$.
The anomalous divergence of the gauge current,
\begin{equation}
{\cal A}^a=(D_M J^M)^a,
\end{equation}
(with the derivative understood to be both gauge-covariant and gravitationally
covariant) is then given by the gauge variation of $W$ as
\begin{equation}
\delta_\alpha W[A]=\int d^4x\int dy\, \sqrt{-g}\,\alpha^a\,{\cal A}^a.
\end{equation}
Since the anomaly is supported only on the branes, we can write
\begin{equation}
\sqrt{-g}\,{\cal A}^a(x,y)={\cal A}^a_1(x)\,\delta(y)+{\cal A}^a_2(x)\,\delta(y-\pi R).
\end{equation}
The ${\cal A}^a_i(x)$ are determined, up to
normalization, by the Wess-Zumino consistency condition. More precisely,
they should be proportional to the usual 4D consistent anomaly:
\begin{equation}\label{4danomaly}
{\cal A}^a_i=\frac{n_i}{24\pi^2}\varepsilon^{\mu\nu\kappa\lambda}\tr\left[T^a\,\partial_\mu \left(A_\nu\partial_\kappa A_\lambda+\frac{1}{2} A_\nu A_\kappa A_\lambda\right)\right].
\end{equation}
Here the gauge fields on the RHS are restricted to the respective branes
(hence we can identify them with their 4D zero modes, up to a common normalization,
 since their bulk profiles are flat).
The constants $n_i$ depend on the number and distribution of the
chiral fermions. Normalizing the trace in Eq.~\eqref{4danomaly} to the 
fundamental representation, a left-chiral fundamental fermion localized on the
$y=0$ ($y=\pi R$) brane will contribute $n_1=1,n_2=0$ ($n_1=0,n_2=1$).
The contribution from a massless left-chiral bulk fermion is $n_1=n_2=1/2$.
On the $S^1/(Z_2\times Z_2')$ orbifold which we are considering, fermions
with orbifold parities $(+-)$ and $(-+)$ may also contribute to localized 
anomalies (even though they do not give rise to 4D zero modes). The 
contribution from a fundamental fermion with parity $(+-)$ is $n_1=-n_2=1/2$;
from a fermion with parity $(-+)$, it is $n_1=-n_2=-1/2$.
Note that, by the topological nature of the anomaly, the 
warp factor never enters here and the discussion is similar to the flat case. 

For the 4D effective theory, obtained by integrating over $y$, to be anomaly-free,
it is necessary and sufficient for  ${\cal A}_1$ and ${\cal A}_2$ to be equal and opposite. 
If ${\cal A}_1=-{\cal A}_2$ and both are nonzero, the 5D theory still
appears anomalous on the branes. In that case, in order to render the theory consistent, 
the anomalous gauge variation of the generating functional
should be compensated by a 5D Chern--Simons term
\begin{equation}
{\cal L}_{\rm CS}=\frac{c}{96\pi^2}\,\varepsilon^{MNOPQ}
\tr\left[A_M F_{NO} F_{PQ}+i\,A_M A_N A_O F_{PQ} -\frac{2}{5} A_M A_N A_O A_P A_Q\right].
\end{equation}
The gauge variation of the Chern--Simons term is a total divergence, leading to equal and opposite
boundary terms on the branes that may precisely cancel the anomaly of Eq.~\eqref{4danomaly} for suitable $c$. 
This is the ``anomaly inflow'' mechanism \cite{Callan:1984sa}.

As an example consider the $\sunf_L$ flavour group of the model in Table \ref{sqcd0-3}. 
There are $\nc$
fundamentals $Q$ on the $y=0$ brane, $\nf$ fundamentals $\eta$ and $\nf$
antifundamentals $\p$ in the bulk (note that $\p^c$ and $\eta^c$ do
not contribute since they have parities $(--)$), and $\nnc$ fundamentals
$q$ as well as $\nf$ antifundamentals $\pp$ on the $y=\pi R$ brane.
The anomalies due to bulk fields cancel; the remaining anomaly is
\begin{equation}\label{lanomaly}
{\cal A}^a_{\sunf_L}=\frac{\nc}{24\pi^2}\varepsilon^{\mu\nu\kappa\lambda}\tr\left[T^a\,\partial_\mu \left(A_\nu\partial_\kappa A_\lambda+\frac{1}{2} A_\nu A_\kappa A_\lambda\right)\right]\;\left[\delta(y)-\delta(y-\pi R)\right]\,e^{4ky}\,.
\end{equation}
It is cancelled by a bulk Chern--Simons term
\begin{equation}
{\cal L}_{\rm CS}=\frac{\nc}{192\pi^2}\,\varepsilon^{MNOPQ}
\tr\left[A_M F_{NO} F_{PQ}+i\,A_M A_N A_O F_{PQ} -\frac{2}{5} A_M A_N A_O A_P A_Q\right].
\end{equation}

From a 4D viewpoint one can therefore simply add the contributions of the zero-modes: the $\sunf_{L}^{3}$ anomaly on the UV brane
is $\nc$, and on the IR brane is $\nnc-\nf=-\nc$, and there is no nett anomaly
contribution from the zero modes of the even bulk fields.
Summarizing the remaining anomalies, the $\Uo_{R}$ charges in Table \ref{sqcd0-3} 
are fixed by the vanishing of the $G^{2}-\Uo_{R}$ anomalies (which implies
$R_{\eta}+R_{\Phi}=2\frac{\nc^{2}}{\nf^{2}}$) and by the $\varphi q\tilde{q}$
and $\eta\varphi$ couplings in the IR superpotentials. The
$\Uo_{B}$ charges are fixed by the $\sunf^{2}-\Uo_{B}$ anomalies.
The gauge group $\nnc=\nf-\nc$ is fixed by the $\sunf^{3}$ anomalies.
Note that $\Uo_{B}$ and $\Uo_{R}$ are global and there remain
uncancelled $\Uo_{R}^{3}$ and $\Uo_{R}-\Uo_{B}^{2}$ anomalies, but these
symmetries are indeed anomaly free with respect to the gauge symmetries
as required.\footnote{Of course also the $\Uo_R^3$ anomaly could be
cancelled, by adding gauge singlets with appropriate $R$-charges.}

\subsection{Baryon matching}

To complete this section we briefly demonstrate another equivalence that can be drawn between the holographic
picture outlined here and that of standard Seiberg duality, namely 
the identification of baryons. In Seiberg duality one of
the major tests of the equivalence of the electric and magnetic formulations
is that the moduli spaces match: in particular there exists a precise
mapping between magnetic and electric baryons. Returning to the SQCD
content of Tables \ref{sqcd0} and \ref{sqcd0-1} for a moment, this
mapping is of the form \begin{equation}
\varepsilon^{(\nf)}\varepsilon^{(\nc)}Q^{\nc}\equiv\varepsilon^{(\nnc)}q^{\nnc},\end{equation}
where the $\varepsilon^{(\nf)}$ Levi-Cevita symbol refers to a contraction
over {}``flavour'' indices (and similar for antibaryons). Note that
the $\varepsilon^{(\nf)}$ yields an object on the LHS that transforms
with $\nf-\nc=\nnc$ antisymmetric flavour indices, exactly matching
the RHS. Indeed this contraction is why the flavour charges on the
quarks have to be negated in the magnetic theory (i.e. the electric
quarks are flavour fundamentals whereas the magnetic ones are flavour
antifundamentals). The baryons can be labelled as follows \begin{equation}
\varepsilon^{(\nf)}B\equiv b\label{eq:ident}\end{equation}
where $B=\varepsilon^{(\nc)}Q^{\nc}$ and $b=\varepsilon^{(\nnc)}q^{\nnc}$
have respectively $\nc$ and $\nnc$ antisymmetric flavour indices.
Now let us return to the AdS model and consider the bulk
$\SU\nfc$ symmetry. Clearly the $\sunc$ and $\sun$ quarks fit into full
$\SU\nfc$ fundamentals so that the quark content in the unbroken theory would be as shown in Table
\ref{tab:Quarks-in-the}. %
\begin{table}
\centering{}\begin{tabular}{|c||c|c|c|c|c|}
\hline 
$ $ & $\SU\nfc$ & $\sunf_{L}$ & $\sunf_{R}$ & $\widetilde{\mathrm U}(1)_{B}$ & $\widetilde{\mathrm U}(1)_{R}$\tabularnewline
\hline
\hline 
$\mathbf{Q}\supset Q,\, q$ & $\fund$ & $\fund$ & 1 & $\frac{1}{\nfc}$ & $0$\tabularnewline
\hline 
$\mathbf{\tilde{Q}}\supset\tilde{Q},\,\tilde{q}$ & $\afund$ & $1$ & $\afund$ & $-\frac{1}{\nfc}$ & $0$\tabularnewline
\hline
\end{tabular}\caption{\emph{Quarks in the unified $\SU\nfc$ theory.\label{tab:Quarks-in-the}}}
\end{table}
The baryons in the unbroken theory would be \begin{equation}
\mathbf{B}=\varepsilon^{(\nfc)}\mathbf{Q}^{\nf},\end{equation}
with $\nfc$ flavour indices. Thus (upto permutation factors)
the singlet object is \begin{eqnarray}
\varepsilon^{(\nf)}\mathbf{B} & = & \varepsilon^{(\nf)}\varepsilon^{(\nc)}\varepsilon^{(\nnc)}Q^{\nc}q^{\nnc}\nonumber \\
 & = & \varepsilon^{(\nf)}B\, b.\end{eqnarray}
This allows us to identify each $B$ with an $\sun$ \emph{anti}quark:
\begin{equation}
\varepsilon^{(\nf)}B\equiv\tilde{b}.\end{equation}
As one might expect the identification is simply the conjugate of
that in Seiberg duality in eq.(\ref{eq:ident}), and indeed the magnetic ``baryon'' $b=q^\nnc$ carries 
baryon charge $-1$.

\section{Supersymmetry breaking}\label{susyb}

Let us now now turn to the question of SUSY breaking. The IR superpotential 
of \eqref{wir} is similar to that of ISS, but with the linear term replaced by a brane 
mass term $\mir\,\hat{\eta}\pp$. Clearly what is required for SUSY breaking 
is a vev for $\eta$. Following the thinking 
outlined in the introduction, we wish to achieve this by a deformation of the UV theory. 
Hence we break the global $R$-symmetry in the UV theory with terms allowed by the 
other symmetries,
\beq
\label{wuv}
\wuv= \mpl \eta \p - \frac{1}{2}\frac{\eta\p\eta\p}{\mpl}\, .
\eeq
A good approximation is to first consider unbroken SUSY on the UV brane. 
These terms can then generate vevs for both $\eta$ and $\p$ as can be seen by setting the $F$-terms equal to zero. 
By a choice of gauge they can be chosen to be diagonal, and then the $D$-term 
contribution to the 
potential sets $\eta = \p = \mpl \mathbbm{1}$ in the vacuum with maximal remaining symmetry.
(Note that the mesons are effectively 
$\nf$ generations of fundamental or antifundamental with respect to a particular $\sunf_L$ or $\sunf_R$.) In this way the flavour symmetry can be broken down to the diagonal symmetry 
$\sunf_L \times\sunf_R \rightarrow \sunf _D$ as in ISS. We may also add a 
\[  \wuv\supset  Q\p \tilde{Q} \]
coupling. Once $\p$ gets a vev, this term generates precisely the $m_Q$ deformation, but with an unsuppressed mass,
 $m_Q\sim \mpl$. 

It is reasonable to assume that some spontaneous breaking of $R$-symmetry 
in the fundamental UV theory generates these terms. The mass term can be generated by 
some $R$-charged singlet fields getting a vev for example, or both it and the quartic 
term may be generated by gravitational effects since the $R$-symmetry is after all only global. Note here a 
marked departure from the ISS model: the $R$-symmetry breaking that 
subsequently appears in the IR theory is under rigid control. 
Indeed according to \eqref{etavev} the vev of $\hat\eta$ is  
\begin{equation}
\hat{\eta}=\mpl\, e^{(\frac{1}{2}-c_{\eta})ky},\end{equation}
so that the effective superpotential \eqref{wir}
becomes the O'Raifeartaigh/ISS one: 
\beq
\label{wir2}
\wir=h \, q\pp \tilde{q} + \missp^2 \pp\, .
\eeq
where \beq \missp^2= e^{-\left( \frac{1}{2}+c_\eta\right) k \pi R}\mpl^2.\eeq
Note that a warped down mass-squared term would be of order 
$e^{-2 k \pi R}\mpl^2$. For $c_\eta=1/2$ the parameter $\missp $ scales with a single 
power of the warp factor because the wave-function of $\hat{\eta}$ is evenly spread over the 
compact dimension. By taking $c_\eta>1/2$ one makes $\missp$ further
exponentially suppressed, due to the localization of the zero-mode of $\hat{\eta}$ on the UV brane.

Hence the IR theory develops a metastable minimum as in ISS \cite{ISS}, and as in ISS it has  a 
remaining anomalous $R$-symmetry responsible for the supersymmetric minima being 
situated at large $\pp$ vev. In order for the 4D description 
to be trustable however there are additional constraints on the parameter $c_\eta$. 
Since the metastable minimum has quark vevs of order $\missp$, and the KK modes of the theory 
have masses of order $\mir=e^{-k\pi R}\mpl$, we have a necessary condition, 
\beq
\missp< \mir\, .
\eeq
This enforces 
\beq c_\eta > \frac{3}{2} \, .\eeq 
Additional constraints arise if one requires the global SUSY minimum (and hence aspects of the 
metastability such as the tunnelling rate) to be well under control  
entirely within the low energy 4D description. A sufficient condition for this would be  
that the $\pp$ vev \eqref{phivev} is less than the KK mode scale $\mir$. 
This depends on the choice of dynamical transmutation scale $\bar{\Lambda}$ in the 
IR theory. In order to have perturbative control this should certainly be greater than $\mir$. 
In the limiting case that $\bar{\Lambda}\sim \mir$ we find the same necessary 
condition $\missp< \mir $, and hence $c_\eta > \frac{3}{2} $,
for the SUSY restoring vacuum to be visible within the 4D theory.
The general necessary condition that includes a valid metastable minimum, 
and a global supersymmetric minimum much further from the origin but less than $\mir$ 
is 
\beq 
\missp \ll \langle \pp \rangle \sim \missp \epsilon^\alpha 
\sim \bar{\Lambda} \epsilon^{\alpha+1} < \mir \ll \bar{\Lambda}\, ,
\eeq 
where $\alpha = \frac{2F-3N}{N}$ and as before $\epsilon = \missp/\bar{\Lambda}$. 
This requires $\alpha$ to be in the range $0> \alpha > -1$, so that for small enough $\epsilon$ 
one can always achieve  
$\missp\ll \langle h\pp \rangle \ll \bar{\Lambda} $. In terms of numbers of colours and 
flavours this becomes 
\beq 
\frac{3 N }{2} > F > N, 
\eeq 
precisely the requirement in the ISS model that one is in the free magnetic phase.  (Of course $F>N$ 
is satisfied in this model by design.)

Since we break $R$ symmetry in the UV, $R$-symmmetry breaking operators 
on the IR brane will generically be induced, and they will lead to additional supersymmetric vacua. 
It is, however, readily checked that these are far away in field space and thus not dangerous. 
First, note that a $\pp^2$ term cannot be induced
even though the remaining symmetry $\sunf_D$ allows it, because the $\sunf_L\times\sunf_R\rightarrow \sunf_D$ 
breaking is spontaneous and proportional to $\vev{\eta}$. Also, since the possible operators would be generated in the UV theory 
(by gravity for example) one would expect them always to be suppressed by powers of $\mpl$. 
(Even if they are absent in the superpotential, there can be similarly suppressed operators in the K\"ahler potential.) 
In this case one can have for instance, a term 
\beq\label{irrbreaking}
\wir~\supset~\frac{(\pp\hat\eta)^2}{\mpl}
\eeq
which would in principle induce a minimum at 
\beq
\langle\pp\rangle\sim\frac{\mir\mpl}{\langle\hat\eta\rangle}\sim\frac{\mir^2\mpl}{(\missp)^2}\gg\mir\,.
\eeq
However, since $\missp\ll \mir$ and $\mpl\gg\mir$ this lies outside the region of validity of the IR theory and 
it cannot destabilize the metastable minimum. (To be very conservative, even if one were to consider the lowest possible
 suppression scale of $\mir$ in Eq.~\eqref{irrbreaking}, the global SUSY restoring minima would still be still sufficiently far away.)

An (appropriately suppressed) IR brane operator of the form $q\Phi\tilde q$ could be useful if 
we identify (part of) the magnetic quarks as messenger fields for gauge mediation. It would effectively 
be a mass term for $q$ and $\tilde q$; since their vevs leave an $\sunc$ subgroup of the $\sunf_D$ 
flavour group unbroken, one could imagine identifying this subgroup with the visible sector gauge 
group. Soft masses would then be generated, as usual in gauge mediation, by $q$ and $\tilde q$ loops.

As a final remark, in the metastable SUSY breaking vacuum the expectation value of the bulk
fields $\eta$ and $\Phi$ (which we determined in a first approximation from the condition for unbroken SUSY) 
will receive small corrections because the potential is now elevated by an additional term $\sim \nc |h \missp^2|^2$. 
These corrections are suppressed by the warping (i.e. they are a correction of order $e^{-2\pi k R} \mpl $ in the vev of $\eta$ on the 
UV brane), so they can be consistently neglected for our analysis.

\subsection{Holographic interpretation}

By the AdS/CFT correspondence \cite{Maldacena:1997re,Gubser:1998bc,Witten:1998qj}, 
an RS1 type model on a ``slice of AdS$_5$'' can be regarded as dual to a 4D conformal 
field theory. The AdS/CFT dictionary relates the building blocks of the 5D theory to 
objects on the CFT side \cite{ArkaniHamed:2000ds,Rattazzi:2000hs}. Specifically, 
the fifth dimension $y$ of AdS$_5$ becomes the renormalization scale in the CFT; the 
truncation of AdS$_5$ at $y=0$ by a UV brane corresponds to introducing a UV cutoff 
scale for the CFT, coupling it to gravity and possibly other fundamental
degrees of freedom; and the truncation
at $y=\pi R$ by an IR brane corresponds to a spontaneous breaking of conformal 
invariance in the infra-red. The CFT is strongly coupled, because gravity on the AdS side 
is weakly coupled (gravitational dynamics is, of course, completely negligible for
our analysis). Bulk gauge groups in AdS correspond to internal global symmetries of 
the CFT which are weakly gauged, so as not to affect the CFT dynamics. Localized 
fields on the UV brane correspond to fundamental, ``elementary'' fields external to 
the CFT, while localized fields on the IR brane are interpreted as ``composite'' bound 
states formed by the CFT degrees of freedom. Fields propagating in the 5D bulk should
be regarded as being partly composite and partly elementary.

It is interesting to see how the 4D CFT interpretation of our model relates to 
a purely four-dimensional, conventional ISS model retrofitted by an additional, strongly
coupled gauge sector. Clearly some aspects are analogous: for instance, by construction 
the far infra-red dynamics is more or less the same. That is,
in the 4D effective theory of our model we can integrate out the $\eta$ as
well as $\Phi$ and the electric quarks at scales below $m_Q$,~$\mpl$. Then we are left with 
only the IR brane degrees of freedom, constituting the magnetic side of an ISS model,
and a pure $\sunc$ gauge theory which is effectively decoupled.

Similarly, the purely elementary degrees of freedom are, according to the AdS/CFT 
dictionary, just those living on the UV brane. In our case these are just
the electric quarks, which together with the $\sunc$ gauge fields form the corresponding 
Seiberg dual electric theory.

In an ISS model, one may think of the small electric quark mass (which is eventually responsible
for dynamical SUSY breaking) as being generated by strong dynamics of an additional gauge
sector (see e.g.~\cite{Dine:2006gm, Dine:2006xt, Aharony:2006my, Brummer:2007ns}).
Our model is similar in the sense that the meson mass term in the magnetic theory
is naturally small, because it is suppressed by the warp factor and by the $\eta$ bulk
profile in the AdS picture. In the CFT picture the reason for its suppression is that 
it is again generated dynamically.

There is, however, an important difference in that strong coupling of the bulk gauge
group never plays a role in our model. This is very much in contrast to the usual
4D picture of Seiberg duality, where the electric gauge group has a Landau pole around
the same scale as the magnetic gauge group, defining where the transition
between electric and magnetic degrees of freedom takes place. In fact, in our model
we are delaying the onset of strong coupling for both the 
magnetic and the electric gauge factors beyond the range of validity of a purely
magnetic or electric description. That is, the $\sun$ and $\sunc$ gauge couplings 
should unify near the compactification scale into the $\mathrm{SU}(\nfc)$ gauge 
coupling, which should be perturbative at that scale. Formally the magnetic sector has a
Landau pole in the UV, which is however at a much higher scale where the description in terms
of far IR degrees of freedom is no longer valid. Likewise, the electric sector would
become strongly coupled in the infra-red if the quark mass was sufficiently small.
However, since $m_Q\sim\mpl$ is large the electric quarks decouple before strong coupling is reached. What 
remains of the electric theory is then a pure SYM theory which couples to the magnetic 
sector only through irrelevant operators.

\subsection*{Acknowledgements} We are grateful to Tony Gherghetta for very useful 
conversations. SAA acknowledges a Leverhulme Research Fellowship.

\end{document}